\documentclass[aps,pra,onecolumn,showpacs,amsmath,amssymb,floatfix,footinbib,superscriptaddress]{revtex4-1}
\usepackage{amsmath,natbib,graphicx,subfigure,amssymb,graphics,amsmath,mathrsfs,CJK,color}
\usepackage{multirow,fancyhdr,color,bm,tabularx,psfrag,dcolumn}
\usepackage[colorlinks=true,linkcolor=blue,urlcolor=blue,citecolor=blue]{hyperref}
\usepackage{tikz,xcolor,hyperref}
\definecolor{lime}{HTML}{A6CE39}
\DeclareRobustCommand{\orcidicon}{%
    \begin{tikzpicture}
    \draw[lime, fill=lime] (0,0)
    circle [radius=0.16]
    node[white] {{\fontfamily{qag}\selectfont \tiny ID}};\draw[white, fill=white] (-0.0625,0.095)
    circle [radius=0.007];
    \end{tikzpicture}
    \hspace{-2mm}}
\foreach \x in {A, ..., Z}
{\expandafter\xdef\csname orcid\x\endcsname{\noexpand\href{https://orcid.org/\csname orcidauthor\x\endcsname}{\noexpand\orcidicon}}}

\begin{document}

\title{\large Photosynthetic properties assisted by the quantum entanglement in two adjacent pigment molecules}

\author{Lu-Xin Xu }
\affiliation{Department of Physics, Faculty of Science, Kunming University of Science and Technology, Kunming, 650500, PR China}

\author{Shun-Cai Zhao\orcidA{}}
\email[Corresponding author: ]{zhaosc@kmust.edu.cn.}
\affiliation{Department of Physics, Faculty of Science, Kunming University of Science and Technology, Kunming, 650500, PR China}

\author{Ling-Fang Li }
\affiliation{Department of Physics, Faculty of Science, Kunming University of Science and Technology, Kunming, 650500, PR China}

\date{\today}

\begin{abstract}
The quantum dynamics of entanglement is widely revealed in photosynthetic light-harvesting complexes. Different from the previous work, we explore the properties of exciton transport and photosynthesis assisted by the quantum entanglement in two adjacent pigment molecules, which are measured by the population dynamics behaviors, the $j$-$V$ characteristics and by the output power via a photosynthetic quantum heat engine (QHE) model. A more robust exciton transport dynamic behavior is compared with those without quantum entanglement, and the photosynthetic characteristics evaluated by the output current and power were proved to be enhanced by the quantum entanglement at different ambient temperatures. These results may point toward the possibility for artificial photosynthetic nanostructures inspired by this quantum biological systems.
\begin{description}
\item[PACs]{42.50.Gy}
\item[Keywords]{Photosynthetic properties; quantum entanglement; two adjacent pigment molecules}
\end{description}
\end{abstract}

\maketitle
\section{Introduction}

Quantum entanglement is a characteristic quantum effect and has been widely explored in recent years\cite{2002Computable,2009Entanglement,N2010Environment,2014Quantum,RevModPhys.81.865,Kokail2021}.
The dynamics of entanglement in generally non-equilibrium quantum systems attracted some interest in several investigations recently\cite{2009Enhanced,2012Nonequilibrium,2016Non,2019Quantum}, and the characteristically quantum signatures of entanglement have been demonstrated in thermal state of bulk systems at different low temperatures \cite{2008Quantifying}. Owing to the fragile property, precisely engineered entangled states is difficult. Hence, indirect access to entanglement is of great interest\cite{Pant2019,Paneru_2020}. Since the entanglement properties can be influenced by the photonic quantum state in different optical mode bases, a constructive method was put forward to obtain families of states remaining entangled for any choice of mode decomposition\cite{PhysRevA.100.062129}. In the quantum information research, the entanglement is used as resource for information processing tasks\cite{2019wang,Sorelli2020,Sharma2021}. In an environment where no static entanglement exists, the entanglement can be continuously generated and destroyed by non-equilibrium effects\cite{Jianming2010Dynamic}.
Recently, it has been recognized that entanglement is a natural feature of coherent evolution\cite{10.1088/1402-4896/abdd55,LingFangLi44215,PhysRevD.103.016007,Belfakir2021,JadotNat,PhysRevA.101.012301}, and some efforts have devoted into the realms where entanglement can exist rigorously, particularly in natural systems not in laboratory conditions\cite{Tiersch2012,Caoeaaz4888}, and entanglement was observed in the complex non-equilibrium biological precess\cite{Caoeaaz4888}.
and establish methods

In a previous significant work\cite{Mohan2010}, the authors established methods for quantifying entanglement by describing necessary and sufficient conditions for entanglement, defined as the relative entropy of entanglement with respect to the set entropy of totally separable states, and concluded that entanglement was still demonstrated in the overdamped case with strong coupling between two sites. However, less work is devoted to the roles of quantum entanglement in photosynthetic properties, such as the exciton transport properties and photosynthetic yield owing to the entanglement between two adjacent pigment molecules. In this work, we introduce an entanglement into two coupled pigment molecules for this proposed target, and present intuitive results by evaluating the population dynamics of photogenerated excitons, the j-V characteristics and power via a photosynthetic quantum heat engine (QHE) model. We begin with the photosynthetic quantum heat engine (QHE) model for two adjacent  pigment molecules, and then deduced the photogenerated excitons transport process of this Born-Markov system. With the goal in mind, we reveal the properties of exciton transport and photosynthetic yield controlled by the quantum entanglement at different ambient temperatures. As the results, we compared between these properties and their counterparts without entanglement, which contributes us to confirm the positive role of quantum entanglement in photosynthesis.

This paper is organized as follows: In Sec. II, we describe the theoretical model for the two adjacent  pigment molecules and study the exciton energy transport process; in Sec. III, we analyze the photosynthesis properties assisted by the quantum entanglement; Sec. IV contains the remarks and conclusions.
the use of logarithmic negativity to characterized entanglement

\section{Theoretical model for the two adjacent pigments}

\subsection{Five-level QHE model for the two adjacent pigments}

\begin{figure}[htp]
\center
\includegraphics[width=0.40\columnwidth]{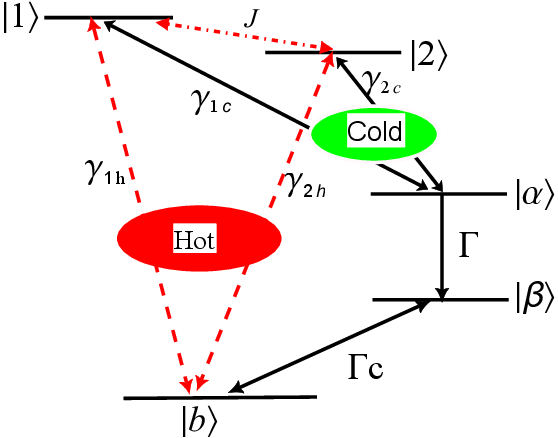}
\caption{(Color online) Corresponding energy-level diagram of the two pigment molecules.}
\label{Fig.1}
\end{figure}

In this work, we will establish a quantum heat engine (QHE) model which has similarly been used to describe double quantum dot photocell systems\cite{PhysRevA.84.053818,PhysRevA.84.053829}, to describe the exciton transport process, and concentrated on the photosynthetic properties assisted by the quantum entanglement between two adjacent pigment molecules. A five-level biologic quantum heat engine (QHE) model is introduced in Fig.\ref{Fig.1}: the states $|1\rangle$ and $|2\rangle$ describe the two excited pigments via the absorption of solar photons, and the sun environment constitutes the Hot bath. The common ground state $|b\rangle$ represents the un-excited two pigments molecules with the lowest energy. Hence, the transitions $|b\rangle$ $\leftrightarrow$ $|1\rangle$ and $|b\rangle$ $\leftrightarrow$ $|2\rangle$ denote the harvesting-light processes in the two adjacent pigment molecules with rates $\gamma_{1h}$ and $\gamma_{2h}$, respectively. The coupling between the two pigment molecules is described by $J$, and the entanglement across the bipartition, i.e., the two adjacent pigments is quantified by using the logarithmic negativity\cite{Plenio2007,PhysRevA.81.062346}. Different from the multi-channel schemes in Refs.\cite{PhysRevA.84.053818,PhysRevA.84.053829,Qin2016A}, the coupling strength $J$ between two pigment molecules in this model, is quantitatively introduced to satisfy the Born-Marco approximation during the derivation of the master equation, i.e., $J$=0.001 (See in the Table \ref{Table.1}), the weak coupling between two pigment molecules. Then the excited electrons are transferred to the acceptors by the emission of a phonon via two pathways: $|1\rangle$ $\leftrightarrow$ $|\alpha\rangle$ and $|2\rangle$ $\leftrightarrow$ $|\alpha\rangle$ at the emissions rates $\gamma_{ic(i=1,2)}$  with $|\alpha\rangle$ being the charge-separated state in the two pigment molecules. And the transition $|\alpha\rangle$ $\leftrightarrow$ $|\beta\rangle$ outputs the photoelectric energy at the rate $\Gamma$ from the two adjacent pigment molecules, the relaxation from $|\beta\rangle$ to the ground state $|b\rangle$ at the rate $\Gamma_{c}$ completes the transport process. Therefore, the transition in the five-level system forms a quantum heat engine (QHE) scheme. With this knowledge, the Hamiltonian for this quantum heat engine (QHE) model is read as,

\begin{align}
 \hat{H}=\hat{H}_{S}+\hat{H}_{B}+\hat{H}_{S-B},
\end{align}
\noindent with
\begin{widetext}\begin{align}
 H_{S}=\sum_{i=1,2}\varepsilon_{i}|i\rangle\langle i|+J(|1\rangle\langle 2|+|2\rangle\langle 1|)+\varepsilon_{\alpha}|\alpha\rangle\langle \alpha|+\varepsilon_{\beta}|\beta\rangle\langle \beta|+\varepsilon_{b}|b\rangle\langle b|
\end{align}\end{widetext}

\noindent $\hat{H}_{S}$ represents the Hamiltonian of the five-level quantum heat engine (QHE) system with the energy $\varepsilon_{i}$ of the ith state, and $J$ denotes the coupling strength between states $|1\rangle$ and $|2\rangle$. $\hat{H}_{B}$ describes the Hamiltonian of the ambient Cold bath with $\hat{a}^{\dag}$($\hat{a}$) being the creation(annihilation) operator and $\omega_{k}$ being the frequency of the kth harmonic oscillator mode, its expression is as follows.

\begin{align}
 \hat{H}_{B}=\sum_{k}\hbar\omega_{k}\hat{a}^{\dag}_{k}\hat{a}_{k},
\end{align}

\noindent The  interaction Hamiltonian $\hat{H}_{S-B}$, i.e., the third part in the Hamiltonian of this quantum heat engine (QHE) model describes the interaction between the system and the ambient Cold bath. Under the rotating-wave approximations\cite{Zhao2019,ZHONG2021104503}, $\hat{H}_{S-B}=\hat{V}_{h}+\hat{V}_{c}$ is given as follows,

\begin{align}
\hat{V}_{h}=\sum_{i=1,2}\sum_{k}\hbar(g_{i,k}^{(h)}\hat{\sigma}_{b,i}\otimes\hat a_{h,k}^{\dag}+g_{i,k}^{(h)\ast}\hat{\sigma}_{b,i}^{\dag}\otimes\hat a_{h,k}),
\end{align}
\begin{widetext}\begin{align}
\hat{V}_{c}=\sum_{k}[\sum_{i=1,2}\hbar(g_{i,k}^{(c)}\hat{\sigma}_{\alpha,i}\otimes\hat a_{c,k}^{\dag}+g_{i,k}^{(c)\ast}\hat{\sigma}_{\alpha,i}^{\dag}\otimes\hat a_{c,k})+\hbar(g_{b,\beta}^{(c)}\hat{\sigma}_{b,\beta}\otimes\hat a_{c,k}^{\dag}+g_{b,\beta}^{(c)\ast}\hat{\sigma}_{b,\beta}^{\dag}\otimes\hat a_{c,k})],
\end{align}\end{widetext}

\noindent where $g_{i,k}^{(h)}$ ($g_{j,k}^{(c)}$) denotes the coupling strength of ith pigment to the kth Hot(Cold) bath mode, $g_{b,\beta}^{(c)}$ describes the coupling strength between the transition $|b\rangle$ $\leftrightarrow$ $|\beta\rangle$ and the kth Cold bath mode. And $a_{h,k}^{\dag}$, $a_{c,k}^{\dag}$($\hat a_{h,k}$, $\hat a_{c,k}$) are the creation (annihilation) operations of the Hot and Cold bath, respectively.
The lowering operators are defined as: $\hat{\sigma}_{b,i}=|b\rangle\langle i|$, $\hat{\sigma}_{b,\beta}=|b\rangle\langle \beta|$, $\hat{\sigma}_{\alpha,i}=|\alpha\rangle\langle i|$.

\subsection{Master equation}

Under the conventional second-order perturbative treatment with respect to $\hat{V}_{h}$ and $\hat{V}_{c}$, and with the Born-Markov approximation and Weisskopf-Wigner approximation, the unitary evolution of the five-level quantum heat engine (QHE) system can be described by the Lindblad master equation\cite{Zhao2019},

\begin{equation}
\frac{d\hat{\rho}}{dt}=-i[\hat{H}_{s},\hat{\rho}]+L_{H}\hat{\rho}+L_{C}\hat{\rho}+L_{\Gamma c}\hat{\rho}+L_{\Gamma}\hat{\rho}
\end{equation}
\noindent The relaxation (with rates $\gamma_{ih}$) process is described by Lindblad super-operators $L_{H}\hat{\rho}$ listed below:
\begin{widetext}
\begin{eqnarray}
L_{H}\hat{\rho}=\sum_{i=1,2}\frac{\gamma_{ih}}{2}[(n_{ih}+1)(\hat{2\sigma}_{b,i}\hat{\rho}\hat{\sigma}_{b,i}^{\dag}-\hat{\sigma}_{b,i}^{\dag}\hat{\sigma}_{b,i}
\hat{\rho}-\hat{\rho}\hat{\sigma}_{b,i}^{\dag}\hat{\sigma}_{b,i}) +n_{ih}(2\hat{\sigma}_{b,i}^{\dag}\hat{\rho}\hat{\sigma}_{b,i}
-\hat{\sigma}_{b,i}\hat{\sigma}_{b,i}^{\dag}\hat{\rho}-\hat{\rho}\hat{\sigma}_{b,i}\hat{\sigma}_{b,i}^{\dag})],\nonumber
\end{eqnarray}
\end{widetext}
\noindent where $\gamma_{ih}$ denotes exciting transition process from $|b\rangle$ to $|i\rangle(_{i=1,2})$, while ${n}_{ih(_{i=1,2})}=[exp\frac{\varepsilon_{ib}}{K_{B}T_{s}}-1]^{-1}$ with the energy difference $\varepsilon_{ib}$ and the sun temperature $T_{s}$. The transport process from two pigments to the output is denoted by the super-operator $L_{C}\hat{\rho}$,

\begin{widetext}\begin{align}
L_{C}\hat{\rho}=\sum_{i=1,2}\frac{\gamma_{ic}}{2}[(n_{ic}+1)(2\hat{\sigma}_{\alpha,i}\hat{\rho}\hat{\sigma}_{\alpha,i}^{\dag}-\hat{\sigma}_{\alpha,i}^{\dag}\hat{\sigma}_{\alpha,i}
\hat{\rho}-\hat{\rho}\hat{\sigma}_{\alpha,i}^{\dag}\hat{\sigma}_{\alpha,i}),
+n_{ic}(2\hat{\sigma}_{\alpha,i}^{\dag}\hat{\rho}\hat{\sigma}_{\alpha,i}-\hat{\sigma}_{\alpha,i}\hat{\sigma}_{\alpha,i}^{\dag}\hat{\rho}-\hat{\rho}\hat{\sigma}_{\alpha,i}\hat{\sigma}_{\alpha,i}^{\dag})],\nonumber
\end{align}\end{widetext}

\noindent which represents the transition $|i\rangle_{(i=1,2)}$$\leftrightarrow$ $|\alpha\rangle$ at the rates $\gamma_{ic}$, and $n_{ic}$ is the photon population: ${n}_{ic}=[exp\frac{\varepsilon_{i\alpha}}{K_{B}T_{a}}-1]^{-1}_{(i=1,2)}$, where $T_{a}$ represents the ambient temperature and $\varepsilon_{i\alpha}$ is the energy difference. Super-operator $L_{\Gamma c}\hat{\rho}$ corresponding to the transition $|\beta\rangle$ $\leftrightarrow$ $|b\rangle$ describes the second low temperature phonon reservoir with the form,

\begin{widetext}\begin{align}
L_{\Gamma_{c}}\hat{\rho}=\frac{\Gamma c}{2}[(N_{c}+1)(2\hat{\sigma}_{b,\beta}\hat{\rho}\hat{\sigma}_{b,\beta}^{\dag}-\hat{\rho}\hat{\sigma}_{b,\beta}^{\dag}\hat{\sigma}_{b,\beta}-\hat{\sigma}_{b,\beta}^{\dag}\hat{\sigma}_{b,\beta}\hat{\rho})
+N_{c}(2\hat{\sigma}_{\beta,b}\hat{\rho}\hat{\sigma}_{\beta,b}^{\dag}-\hat{\rho}\hat{\sigma}_{\beta,b}^{\dag}\hat{\sigma}_{\beta,b}-\hat{\sigma}_{\beta,b}^{\dag}\hat{\sigma}_{\beta,b}\hat{\rho})],\nonumber
\end{align}\end{widetext}

\noindent with the occupation number $N_{c}=[exp\frac{\varepsilon_{b\beta}}{K_{B}T_{a}}-1]^{-1}$ and energy difference $\varepsilon_{b\beta}$. $L_{\Gamma}\hat{\rho}$ describes a process that the system in state $|\alpha\rangle$ decays to state $|\beta\rangle$, which leads to the current proportional to the relaxation rate $\Gamma$ as defined later,

\begin{align}
L_{\Gamma}\hat{\rho}=\frac{\Gamma}{2}[2\hat{\sigma}_{\beta\alpha}\hat{\rho}\hat{\sigma}_{\beta\alpha}^{\dag}-\hat{\rho}\hat{\sigma}_{\beta\alpha}^{\dag}\hat{\sigma}_{\beta\alpha}-\hat{\sigma}_{\beta\alpha}^{\dag}\hat{\sigma}_{\beta\alpha}\hat{\rho}],\nonumber
\end{align}

Invoking the Weisskopf-Wigner approximation and in the Schrodinger picture, we can get the dynamic equations of the corresponding reduced matrix elements as follows:

\begin{widetext}\begin{eqnarray}
&&\dot{\rho}_{11}=-\gamma_{1h}[(n_{1h}+1)\rho_{11}-n_{1h}\rho_{bb}]-\gamma_{1c}[(n_{1c}+1)\rho_{11}-n_{1c}\rho_{\alpha\alpha}]+iJ(\rho_{12}-\rho_{21}),\nonumber\\
&&\dot{\rho}_{22}= -\gamma_{2h}[(n_{2h}+1)\rho_{22}-n_{2h}\rho_{bb}]-\gamma_{2c}[(n_{2c}+1)\rho_{11}-n_{2c}\rho_{\alpha\alpha}]-iJ(\rho_{12}-\rho_{21}),\nonumber\\
&&\dot{\rho}_{\alpha\alpha}=\gamma_{1c}[(n_{1c}+1)\rho_{11}-n_{1c}\rho_{\alpha\alpha}]+\gamma_{2c}[(n_{2c}+1)\rho_{22}-n_{2c}\rho_{\alpha\alpha}]-\Gamma\rho_{\alpha\alpha},\\
&&\dot{\rho}_{\beta\beta}=\Gamma\rho_{\alpha\alpha}-\Gamma_{c}(N_{c}+1)\rho_{\beta\beta}+\Gamma_{c}N_{c}\rho_{bb},\nonumber\\
&&\dot{\rho}_{12}=i\rho_{12}\varepsilon_{12}-iJ(\rho_{22}-\rho_{11})-\frac{\gamma_{1h}}{2}(n_{1h}+1)\rho_{12}-\frac{\gamma_{2h}}{2}(n_{2h}+1)\rho_{12}-\frac{\gamma_{1c}}{2}(n_{1c}+1)\rho_{12}-\frac{\gamma_{2h}}{2}(n_{2c}+1)\rho_{12}.\nonumber
\end{eqnarray}\end{widetext}

\noindent where $\rho_{ii}$ represent the diagonal element and $\rho_{ij}$ describe the non-diagonal element of the corresponding state. In this proposed quantum heat engine (QHE) model, we will introduce the logarithmic negativity\cite{2009Entanglement,Plenio2007,PhysRevA.81.062346} to quantify the intensity of quantum entanglement generated by the two adjacent pigment molecules. Hence, the entanglement between states $|1\rangle$ and $|2\rangle$
is written as follows,

\begin{eqnarray}
A(1|2)=\log_{2}(1-\rho_{bb}+\sqrt{\rho^{2}_{bb}+4|\rho_{12}|^{2}})
\end{eqnarray}

\noindent which allows us to quantitatively analyze the entanglement across a bipartition of a composite system. If the coherence is vanishing, i.e. $\rho_{12}$= 0, there is no entanglement across any partition in the one-excitation sector. It is an accepted and the same computable entanglement monotone\cite{Plenio2007} for arbitrary local operations and arbitrary excitation levels. In order to evaluate quantitatively the exciton energy transport, we introduce an effective voltage $V$ as a drop of the load to the transition $|\alpha\rangle$ $\leftrightarrow$ $|\beta\rangle$, we obtain $eV =\varepsilon_{\alpha\beta}+K_{B}T_{a}\ln\frac{\rho_{\alpha\alpha}}{\rho_{\beta\beta}}$, which defines the voltage $V$ with $e$ being the elementary electric charge. Therefore, the electron is released out of the system with the relaxation rate $\Gamma$, which results in a current from $|\alpha\rangle$ to $|\beta\rangle$ driving a chain of chemical reactions, eventually leads to the stable storage of solar energy. The current is decided by the relaxation rate $\Gamma$ and the population$\rho_{\alpha\alpha}$ with the following expression\cite{ZHAO2020106329,ZHONG2021104094},
\begin{equation}
 j=e\Gamma\rho_{\alpha\alpha}
\end{equation}
Hence, the output power of the proposed five-level quantum heat engine (QHE) system can be calculated as follows\cite{PhysRevB.99.075433,2020Different},
\begin{equation}
 P=jV
\end{equation}

\begin{table}
\begin{center}
\caption{Parameters used in the numerical simulation.}
\label{Table.1}
\vskip 0.2cm\setlength{\tabcolsep}{0.5cm}
\begin{tabular}{ccc}
\hline
\hline
                                                         & Values                 & Units \\
\hline
\(\varepsilon_{12}\)                                     & 0.02                    & eV    \\
\(\varepsilon_{1\alpha}\)                                & 0.2                    & eV    \\
\(\varepsilon_{2\alpha}\)                                & 0.2                    & eV    \\
\(\varepsilon_{ \alpha\beta}\)                           & 0.3                    & eV    \\
\(\varepsilon_{ b\beta}\)                                & 0.2                    & eV    \\
\(\gamma_{1h}\)                                          & 4.2                    & eV    \\
\(\gamma_{2h}\)                                          & 1.98                   & eV    \\
\(\gamma_{1c}\)                                         & 0.012                & eV    \\
\(\gamma_{2c}\)                                         & 0.012                & eV    \\
\(\Gamma\)                                              & 0.12                 & eV    \\
\(\Gamma_{C}\)                                          & 0.024                & eV    \\
\(J\)                                                   & 0.001                &       \\
\(n_{1h}\)                                              & 250                  &       \\
\(n_{2h}\)                                              & 280                  &       \\
\hline
\hline
\end{tabular}
\end{center}
\end{table}

\section{Results and discussions}

\begin{figure}[htp]
\center
\includegraphics[width=0.48\columnwidth]{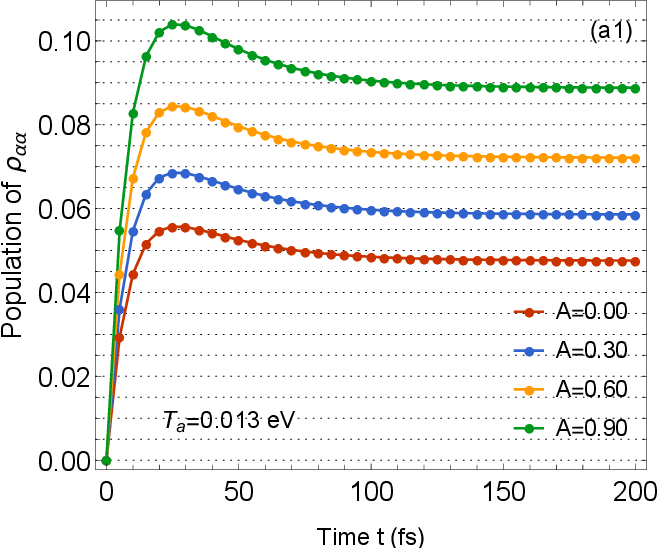}\includegraphics[width=0.48\columnwidth]{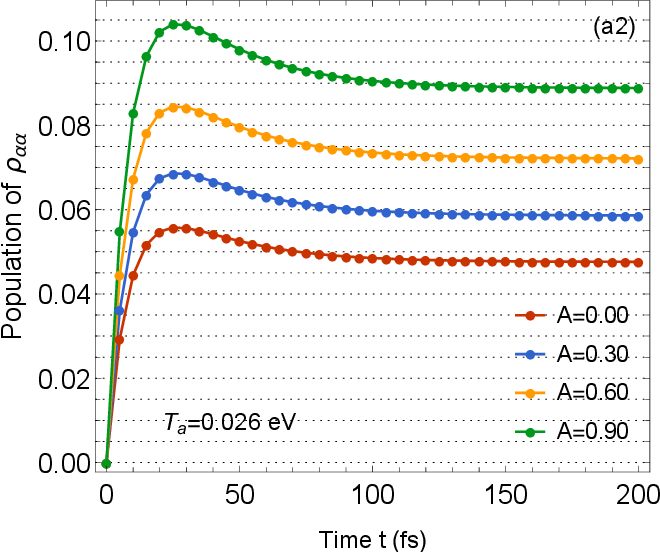}
\hspace{0in}%
\includegraphics[width=0.48\columnwidth]{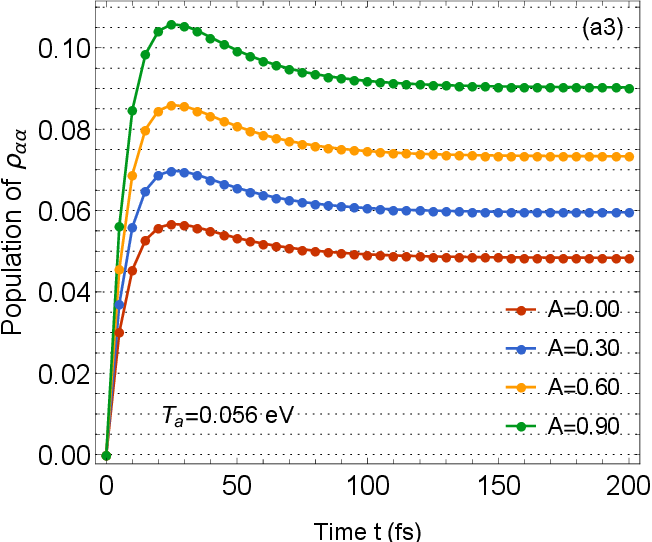 }\includegraphics[width=0.48\columnwidth]{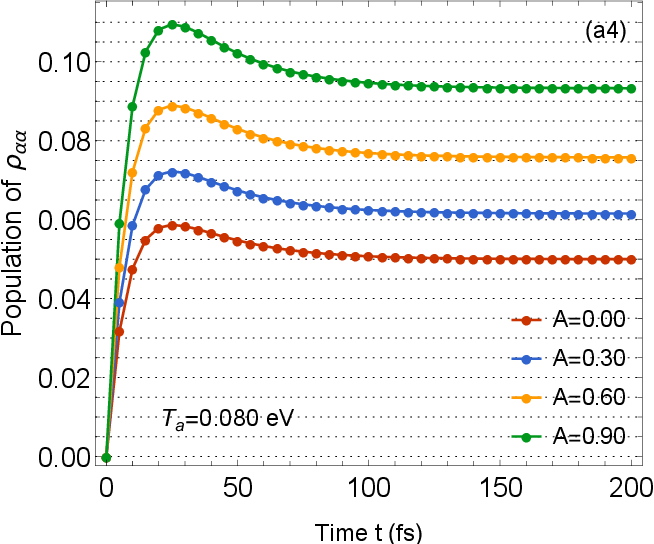 }
\caption{(Color online) Population evolution of $\rho_{\alpha\alpha}$ assisted by the different quantum entanglement intensities $A$ at different ambient temperatures. (a1) $T_{a}$=0.013 $eV$, (a2) $T_{a}$=$0.026 eV$, (a3) $T_{a}$=$0.056 eV$, (a4) $T_{a}$$=$0.080 $eV$. All the other parameters are selected from in Table \ref{Table.1}.}
\label{Fig.2}
\end{figure}

In order to quantitatively evaluate the assists of quantum entanglement to exciton transport and photosynthetic performances, the parameters used in this work are listed detailedly in Table \ref{Table.1}. In this  proposed theoretical model, the dynamic behaviors of population on the charge-separated state $|\alpha\rangle$ [in Fig.\ref{Fig.1}] play the key role in the process of exciton transport. Therefore, Fig.\ref{Fig.2} plots the population evolution of $\rho_{\alpha\alpha}$ regulated by the entanglement intensities $A$ at different ambient temperatures. The curves in Fig.\ref{Fig.2} shows two distinctive features: one is that the populations are greatly enhanced by the intensities of entanglement $A$, and the peak populations increase with the entanglement intensities $A$. Especially, the stable peak populations are much larger with $A$=0.9 than those without quantum entanglement (i.e., $A$=0.0), and the green and red curves in Fig.\ref{Fig.2} demonstrate the results. The other is that the room temperature ($T_{a}$=$0.026 eV$) is the divide of the peak populations on the charge-separated state $|\alpha\rangle$. When $T_{a}$ is less than or equal to $0.026 eV$, the peak populations are almost constant corresponding to the same $A$. For example, when $A$ =0.6, the peak value of $\rho_{\alpha\alpha}$ is about 0.085 with the ambient temperatures $T_{a}$=0.013 $eV$ and $0.026 eV$, respectively. However, when $T_{a}$ is greater than $0.026 eV$, the peak populations are larger and increase with $T_{a}$. When $A$=0.9 with $T_{a}$=0.056 eV and 0.080 eV in Fig.\ref{Fig.2}(a3) and (a4), we notice that the peak populations are 0.0105 and 0.011, respectively. These results indicate that more excitons were transited to the state $|\alpha\rangle$ due to population instability on states $|1\rangle$ and $|2\rangle$ caused by increasing $A$. Therefore, the room temperature plays a key role in the efficient exciton transport between two adjacent pigment molecules. Further revealing the quantum kinetic realm of ambient temperature on the exciton transport is attractive and the research content in our next phase.

\begin{figure}[htp]
\center
\includegraphics[width=0.48\columnwidth]{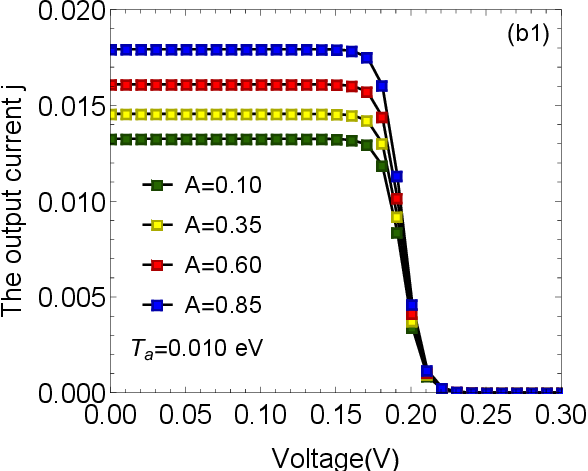}\includegraphics[width=0.48\columnwidth]{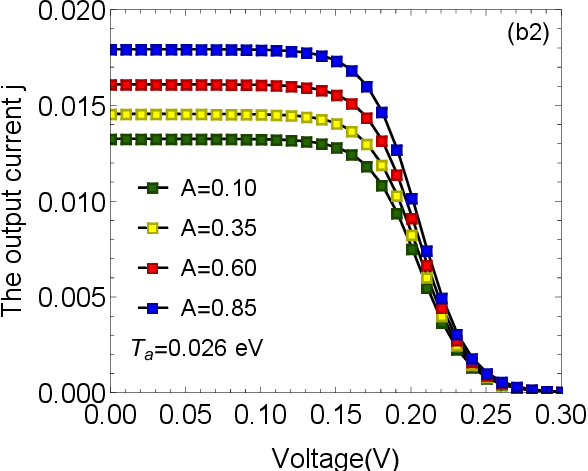}
\hspace{0in}%
\includegraphics[width=0.48\columnwidth]{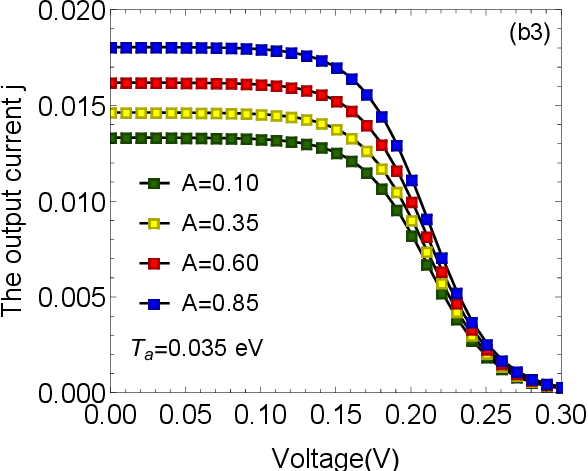}\includegraphics[width=0.48\columnwidth]{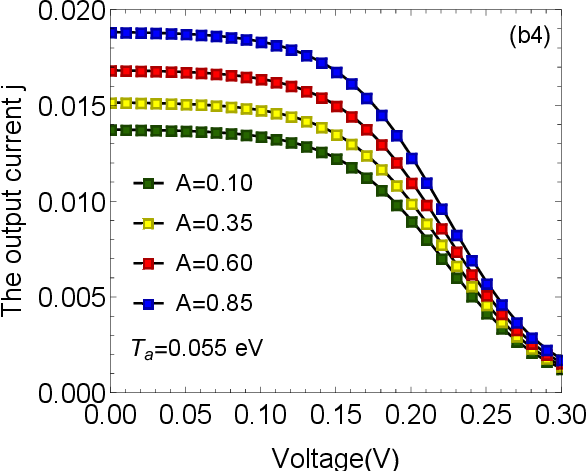}
\caption{(Color online) Current-voltage characteristics $j$-$V$ assisted by the different quantum entanglement intensities $A$ at different temperatures. (b1) $T_{a}$$=$0.010 $eV$, (b2) $T_{a}$$=$0.026 $eV$, (b3) $T_{a}$$=$0.035 $eV$, (b4) $T_{a}$$=$0.055 $eV$. All the other parameters are selected from in Table \ref{Table.1}.}
\label{Fig.3}
\end{figure}

The positive influence of entanglement on the exciton transport was shown clearly by their dynamic behaviors. Meanwhile, the effect of entanglement on the steady photosynthetic performances is much attractive at different ambient temperatures. Fig.\ref{Fig.3} reveals the steady photosynthetic features regulated by the entanglement $A$ at different ambient temperatures. It is well known that the fill factor (FF) is a key indicator to measure the performance of photovoltaic cells\cite{HOFFLER2016180,Chiang2016,Surdo201701190}. Fig.\ref{Fig.3} plots the $j$-$V$ characteristics assisted by the different quantum entanglement intensities $A$ and by different ambient temperatures. As the curves shown in Fig.\ref{Fig.3}, the short-cut currents are increasing from $A$=0.10 to 0.85 and the open-circuit voltages are increasing with the increment of $T_{a}$ from (b1) to (b4), but FF decreases with the increment of ambient temperatures. It indicates that the photosynthetic performances of the proposed model can be enhanced by the entanglement $A$ while be inhibited by the ambient temperatures. We infer that the physical mechanism of $j$-$V$ characteristics in Fig.\ref{Fig.3} is consistent with the behavior of population on the state $|\alpha\rangle$. Owing to the increasing entanglement, the greater population probability in state $|\alpha\rangle$ results in stronger exciton transport, which leads to the greater short-circuit currents in Fig.\ref{Fig.3}. In the meantime, some negative factors, such as the radiative recombination\cite{Zhaoch2020} can be enhanced by the increasing ambient temperatures. Therefore, these mentioned factors lead to the gradual decrease of FF in Fig.\ref{Fig.3}.

\begin{figure}[htp]
\center
\includegraphics[width=0.48\columnwidth]{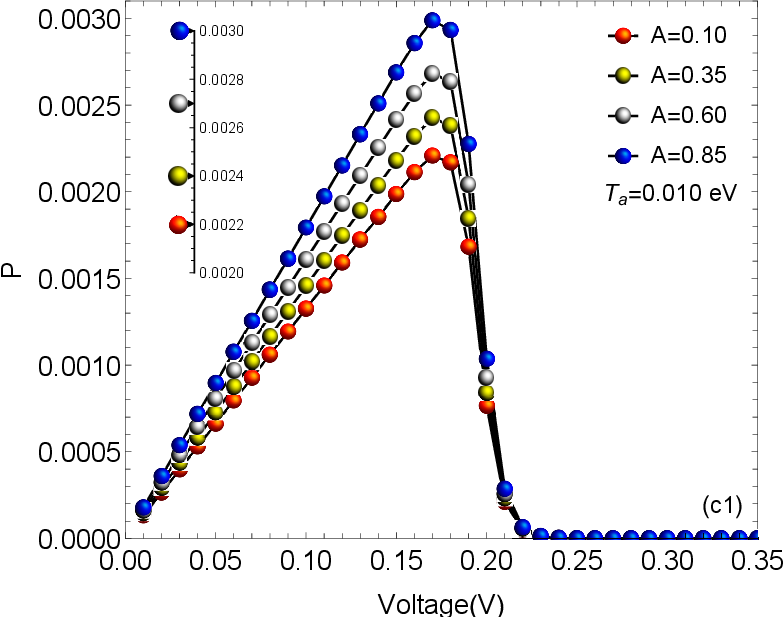}\includegraphics[width=0.48\columnwidth]{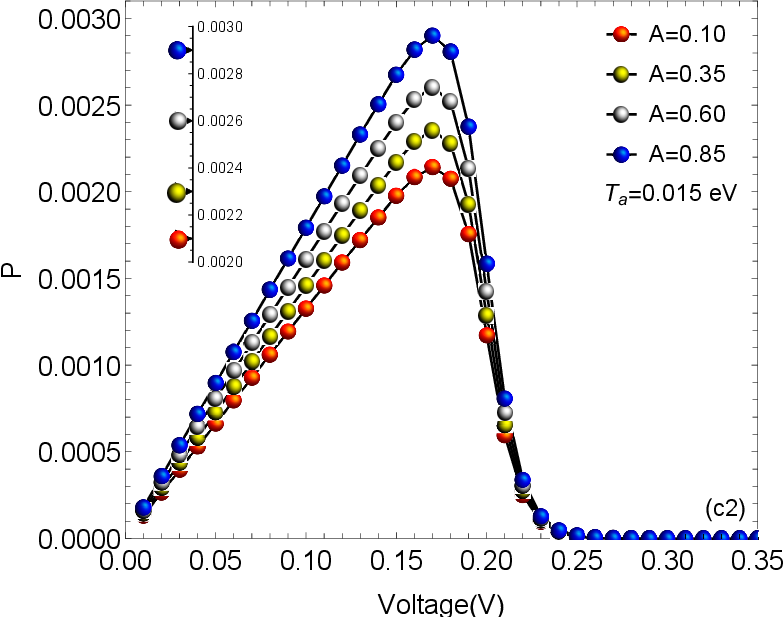}
\hspace{0in}%
\includegraphics[width=0.48\columnwidth]{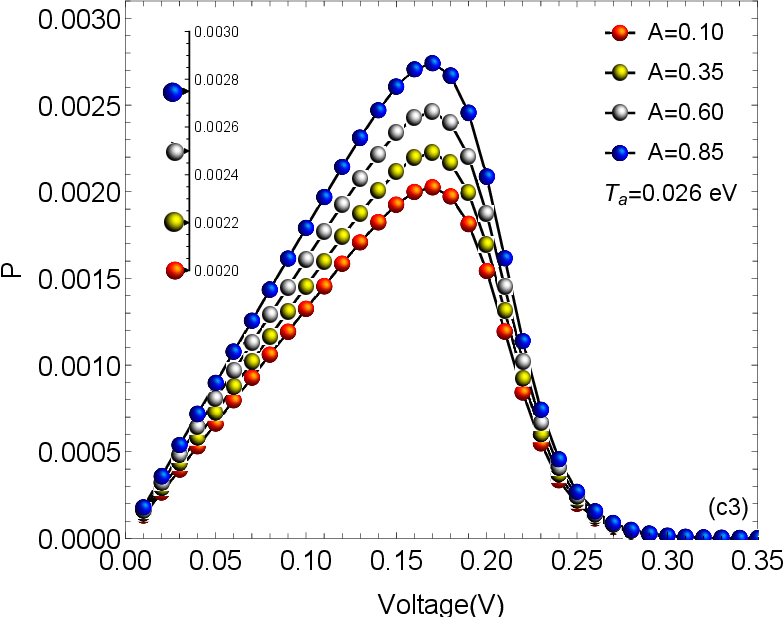}\includegraphics[width=0.48\columnwidth]{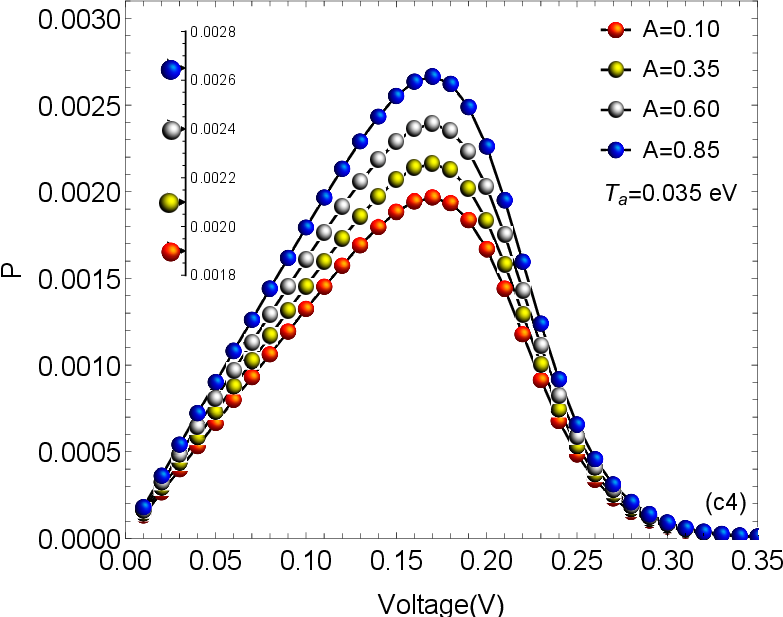}
\caption{(Color online)Power $P$ versus the output voltage $V$ assisted by the different quantum entanglement intensities $A$ at different temperatures.  (c1) $T_{a}$$=$0.010 $eV$, (c2) $T_{a}$$=$0.015 $eV$, (c3) $T_{a}$$=$0.026 $eV$, (c4) $T_{a}$$=$0.035 $eV$. The inner illustrations show their peak powers. All the other parameters are selected from in Table \ref{Table.1}.}
\label{Fig.4}
\end{figure}

The gradual decrease of FF in Fig.\ref{Fig.3} may portend some passive influence on the output power of the two adjacent pigment molecules, and the curves in Fig.\ref{Fig.4} demonstrate the prediction. As shown in Fig.\ref{Fig.4}, the output power $P$ versus the voltage $V$ are shown at different ambient temperatures. It notes that the peak power increases with $A$ about at $V$=0.17 V with the same ambient temperature. However, the increasing $T_{a}$ suppresses the peak powers when $A$ takes the same value, and the peak values shown by the inner illustrations prove this conclusion in Fig.\ref{Fig.4} from (c1) to (c4). The above results indicate that the proper regulation between the intensity of quantum entanglement and ambient temperature is a key issue for efficient photosynthetic performance in this proposed photosynthetic system.

Finally, this work has contributed this effort with the demonstration from theoretical methodology of optimal control on the photosynthesis properties via the entanglement at different ambient temperatures. Although these mentioned results need further experimental confirmation, some potential optimal strategies are provided for artificial bionic photovoltaic devices at proper ambient temperature. In future work, based on the techniques presented here, we are planning to consider more general non-Markovian models and other biomolecular complexes, and we will investigate the importance of relation between multi-exciton entanglement and ambient temperature. We will also explore the use of quantify the intensity of quantum entanglement, and explain the realm between the entanglement and ambient temperature in this context.

\section{Conclusion}

In conclusion, we discuss the role of quantum entanglement in the exciton transport and photosynthetic features of two adjacent pigment molecules via the quantum heat engine (QHE) model at different temperatures. Different from previous work\cite{Scully2011Quantum,Dorfman2013Photosynthetic}, we are not concerned with the description of the construction of quantum heat engine (QHE) but with the extraction of its physical significance, nor with the difference of coupling strength between two adjacent pigment molecules. Instead, we consider Born-Markov approximation under the weak coupling coefficient $J$ between two pigment molecules. And the results demonstrate that the enhanced population evolution, steady-state $j$-$V$ and $P$-$V$ characteristics can be assisted by the quantum entanglement at proper ambient temperatures. Unlike previous work\cite{2014Quantum,2009Enhanced,Jianming2010Dynamic,PhysRevA.81.062346} which discussed the character of quantum entanglement in different biological systems, this work demonstrates that photosynthetic power and exciton transport can be promoted by quantum entanglement. Firstly, the results may provide some theoretical strategies for artificial photosynthetic devices hinted from the entanglement. Secondly, this work describe the photosynthetic system via the quantum heat engine (QHE) model, which provides a theoretical methodology for quantum biology.

\section*{Competing Interests}

The authors declare no competing financial or non-financial interests. This article does not contain any studies with human participants or animals performed by any of the authors. Informed consent was obtained from all individual participants included in the study.

\section*{Author contributions}

S. C. Zhao conceived the idea. L. X. Xu performed the numerical computations and wrote the draft, and S. C. Zhao did the analysis and revised the paper. L. X. Xu and L. F. Li participated in part of the discussion.

\section*{Data Availability Statement}

This manuscript has associated data in a data repository.[Authors' comment: All data included in this manuscript are available upon resonable request by contaicting with the corresponding author.]

\section*{Acknowledgments}

This work is supported by the National Natural Science Foundation of China ( Grant Nos. 62065009 and 61565008 ), Yunnan Fundamental Research Projects, China( Grant No. 2016FB009 ).

\bibliography{reference}
\bibliographystyle{unsrt}
\end{document}